# Pairing Symmetry, Spin-Gap and more in HTSC Cuprates.

Moshe Dayan

Department of Physics, Ben-Gurion University,

Beer-Sheva 84105, Israel.

# Pairing Symmetry, Spin-Gap and more in HTSC Cuprates.

Moshe Dayan

Department of Physics, Ben-Gurion University,

Beer-Sheva 84105, Israel.

## Abstract

This work is based on the double correlated linear aggregations of holes in checkerboard geometry. It is proved that the pairing function symmetry is $-d_{x^2-y^2}$, as been observed experimentally. It is also shown that there is a "superconductive spin gap" for the observation of the magnetic incommensurate modulation peaks, in agreement with experiment. In addition, the unperturbed Hamiltonian and its related propagator are reanalyzed and modified.

1. **INTRODUCTION**

In recent two papers by the present author the linear aggregation of holes in checkerboard geometry has been established in HTSC cuprates [1,2]. In one of these papers, fluctuating deviations from full columns or rows have been attributed to superconductive pairing [2]. This aggregation of holes may be considered as a spontaneous symmetry breaking, from two dimensional, to a product of two one dimensional symmetries, which hereafter is referred to as semi one dimensional. It has been demonstrated that this semi one dimensional symmetry has a profound effect on the (so called) "Fermi surfaces" of the cuprates. Moreover, it induces internal spin density waves (SDW) that have been observed by Neutron scattering measurements (NSM), and checkerboard $4a \times 4a$ structures that have been observed by STM. While the latter kind of structures has been discussed only qualitatively in [1], the basic existence of the static form of the former one has been shown quantitatively.

There are many subjects that have been discussed only partially, and their further investigation is in the waiting line. Such is the effect of the fluctuating strings of holes on many relevant physical functions. These fluctuations, combined with the reduced dimensionality should presumably have profound effects on many physical functions, and on their temperature dependence. An important matter which has not been approached is the finite temperature treatment, as the treatment so far is a zero temperature one. Another issue which has hardly been investigated is the various interactions that are responsible to the pseudogap and to superconductivity. In this respect, one should acknowledge the magnetic interaction that has been proposed and discussed in the last paper [2], but still that discussion is preliminary. Electron transport properties should also be investigated within the framework of the linear aggregation of holes. Another subject that requires attention is the dispersion of the Neutron scattering peaks that modulate the $2\pi a^{-1}(0.5, 0.5)$ anti-ferromagnetic (AFM) wavevector, and are shown as the satellites $2\pi a^{-1}(0.5 \pm \delta, 0.5)$, and $2\pi a^{-1}(0.5, 0.5 \pm \delta)$, where $\delta$ is roughly the doping. In the present paper I refrain from dealing with such subjects.

The first subject to be addressed in the present paper is the symmetry of the superconductive pairing state. There is enough experimental evidence to suggest that the pairing state function is mainly of the $d_{x^2-y^2}$ symmetry [3-6], which is intuitively compatible with the checkerboard geometry. Several kinds of experimental results reflect the checkerboard semi one dimensional symmetry in the cuprates. Such are the ARPES results that provide the shapes of the "Fermi surfaces" in the reciprocal space, the checkerboard $4a \times 4a$ structures observed by STM, and the results of NSM. However, the most convincing measurements which suggest specifically the $d_{x^2-y^2}$ symmetry for the pairing state are phase sensitive squid measurements [3,4]. Experiments by the double-junction corner squid configuration measure directly the pairing phase difference between the x and the y directions, thus proving directly the $d_{x^2-y^2}$ symmetry [5,6]. Intuitively, our model of linear aggregation of holes in two directions seems to suggest naturally the $d_{x^2-y^2}$ symmetry, or at least some kind of one dimensional symmetry, in two directions. However, some mathematical effort is needed to establish this intuition rigorously. This is done in the present paper.

An important observed aspect of the incommensurate Neutron scattering peaks is its lack of mutual appearance with superconductivity, a phenomenon known as the spin-gap [7-10]. Experimental results show that, below $T_c$, the Neutron satellite peaks are absent in the energy range of the superconductive gap, whereas above $T_c$, they exist down to zero energy [7-10]. This suggests their impossible coexistence with the superconductive symmetry. Here we prove that this conclusion is in a complete agreement with our model.

Finally, we re-examine our previous analysis regarding the unperturbed Hamiltonian- $H_0$. A careful examination of $H_0$ in the former two papers shows some disagreement between the two. It turns out that the source of the disagreement stems from the Fourier transforms of the basic excitation components of the vector $\hat{O}_k$ in Eq.(21) of Ref.[2]. We notice that each of these excitation components is a linear combination of operators with different wavenumbers. Thus, the excitation components have internal wavenumber structure, which is lost in a regular Fourier transformation. Therefore, when attempting the derivation of $H_0$, one should exercise the scheme of Eqs.(12a,b) in [2], where the wavenumber of each basic operator is taken into account. This is done in the present paper for the two and the four dimensional Hamiltonians. One of our criterions is that the 4-dimensional $H_0$ converges to two 2-dimensinal $H_0$, when superconductivity approaches zero. This is actually achieved in the following.

## 2. THE SYMMETRY OF THE SUPERCONDUCTIVE PAIRING.

We start this analysis with Eq.(14) of Ref. [2], keeping the same nomenclature

$$|\psi_k> = \sum_{i=0}^{N/2} (\frac{u_k}{v_k})^i \sum_{b_i,b_j} [(c^{\pm}_{\bar{k}_{1\uparrow}} c^+_{-\bar{k}_{1\downarrow}} c^+_{k_{1\uparrow}} c^+_{-k_{1\downarrow}} ... c^{\pm}_{\bar{k}_{j\uparrow}} c^+_{-\bar{k}_{j\downarrow}} c^+_{k_{j\uparrow}} c^+_{-k_{j\downarrow}})_{b_j}]_{b_i} C^+_{\pm k \uparrow \downarrow}(b_i) \qquad (1a)$$

$$C^+_{\pm k \uparrow \downarrow}(b_i) = v_k^{N/2} \sum_{n,m=0}^{N/2-i} (\frac{w_k}{v_k})^n \sum_{b_n \neq b_m \neq b_l \neq b_i} [(c^{\pm}_{\bar{k}_{1\uparrow}} c^+_{-k_{1\downarrow}} + c^+_{k_{1\uparrow}} c^+_{-\bar{k}_{1\downarrow}})...(c^{\pm}_{\bar{k}_{n\uparrow}} c^+_{-k_{n\downarrow}} + c^+_{k_{n\uparrow}} c^+_{-\bar{k}_{n\downarrow}})]_{b_n}$$

$$\times (\frac{\theta_k}{v_k})^m [c^+_{\bar{k}_{1\uparrow}} c^+_{-\bar{k}_{1\downarrow}} ... c^+_{\bar{k}_{m\uparrow}} c^+_{-\bar{k}_{m\downarrow}}]_{b_m} [c^+_{k_{1\uparrow}} c^+_{-k_{1\downarrow}} ... c^+_{k_{l\uparrow}} c^+_{-k_{l\downarrow}}]_{b_l} |0> \qquad (1b)$$

It is a weighted sum of combinations $\{b_j\}$, where each combination is a product of terms, each including four operators product of the sort $c^+_{k_{\mu\uparrow}} c^+_{-k_{\mu\downarrow}} c^+_{\bar{k}_{\mu\uparrow}} c^+_{-\bar{k}_{\mu\downarrow}}$. In order to examine the spatial symmetry, one should transform $|\psi_k>$ to exhibit its spatial dependence. One should also be reminded that the checkerboard geometry assumes that the state function is a product between equivalent columns and rows functions. In order to reduce confusion we use different notations for columns and rows. Spatial vectors and wavevectors relevant for columns of holes are denoted by $r = x\hat{x} + y\hat{y}$ and $k = k_x\hat{x} + k_y\hat{y}$, respectively. The quantities $\hat{x}$, and $\hat{y}$ are unit vectors in the x, and y directions, respectively. Spatial vectors and wavevectors relevant for rows of holes are denoted by $\rho = \rho_x\hat{x} + \rho_y\hat{y}$ and $q = q_x\hat{x} + q_y\hat{y}$, respectively. Let's also denote: $r_{12} = r_1 - r_2$, and $\rho_{12} = \rho_1 - \rho_2$. We write

$$\prod_{\mu(b_i,b_j)} ((c^{\pm}_{\bar{k}_{\mu\uparrow}} c^+_{-\bar{k}_{\mu\downarrow}} c^+_{k_{\mu\uparrow}} c^+_{-k_{\mu\downarrow}})_{b_j})_{b_i} = \prod_{\mu(b_i,b_j)} A_\mu \sum_{r_{\mu,12}} [\exp(ik_\mu r_{\mu,12}) c^+_{\mu,r_1\uparrow} c^+_{\mu,r_2\downarrow}] \sum_{\bar{r}_{\mu,12}} [\exp(i\bar{k}_\mu \bar{r}_{\mu,12}) c^+_{\mu\bar{r}_1\uparrow} c^+_{\mu\bar{r}_2\downarrow}]$$

$$= B(b_j,b_i) \prod_{k_\mu} \sum_{r_{\mu,12}, \bar{r}_{\mu,12}} \exp i(k_\mu r_{\mu,12} + \bar{k}_\mu \bar{r}_{\mu,12}) c^+_{r_{\mu,12}} c^+_{\bar{r}_{\mu,12}} \qquad (2)$$

In Eq. (2) $A_\mu$ and $B(b_j,b_i)$ are constants, $c^+_{r_{\mu,12}} = c^+_{\mu r_1 \uparrow} c^+_{\mu r_2 \downarrow}$, and $c^+_{\bar{r}_{\mu,12}} = c^+_{\mu \bar{r}_1 \uparrow} c^+_{\mu \bar{r}_2 \downarrow}$. It has been shown that $c^+_{k_\mu}$ and $c^+_{\bar{k}_\mu}$ are Fourier transformations of the same column (using different exponential Fourier factors) [2]. They are also defined to keep magnetic coherence in accordance with Eq. (7) and Eq. (8) of [2]. The wavevectors $k_\mu$ are denoted according to their transversal $\mu$ components, although their longitudinal components are included implicitly, and are independent of $\mu$. The vector $r_{\mu,12}$ (and $\bar{r}_{\mu,12}$) is spanned between the two members of a Cooper-pair which are located on two different columns. When the left hand side of Eq. (2) is multiplied by an equivalent row function we get

$$\prod_{\mu(b_i,b_j),\nu(a_m,a_n)} ((c^+_{\bar{k}_\mu\uparrow} c^+_{-\bar{k}_\mu\downarrow} c^+_{k_\mu\uparrow} c^+_{-k_\mu\downarrow})_{b_j})_{b_i} ((c^+_{\bar{q}_\nu\uparrow} c^+_{-\bar{q}_\nu\downarrow} c^+_{q_\nu\uparrow} c^+_{-q_\nu\downarrow})_{a_n})_{a_m} =$$

$$= C_{ijmn} \prod_{k_\mu,q_\nu} \sum_{r_{\mu,12},\bar{r}_{\mu,12}} \sum_{\rho_{\nu,34},\bar{\rho}_{\nu,34}} \exp[i(k_\mu r_{\mu,12} + \bar{k}_\mu \bar{r}_{\mu,12} + q_\nu \rho_{\nu,34} + \bar{q}_\nu \bar{\rho}_{\nu,34})] c^+_{r_{\mu,12}} c^+_{\bar{r}_{\mu,12}} c^+_{\rho_{\nu,34}} c^+_{\bar{\rho}_{\nu,34}}$$

(3)

Now we can write $|\psi_k \psi_q>$. In accordance with [1,2], we define $|\psi_k \psi_q>$ as a normalized sum of Eq. (1) over the two time reversal AFM phases- A and B. Here we refer to the two operators $a^+_{k\uparrow}$ and $a^+_{-k\downarrow}$ as time reversal as is the costume in the superconductive literature (although the time reversal of $a^+_{k\uparrow}$ is actually $a_{-k\downarrow}$). Thus, we get

$$|\psi_k \psi_q> = \sum_{i,m=0}^{N/2} (\frac{u_k}{v_k})^i (\frac{u_q}{v_q})^m \sum_{b_i,b_j,a_m,a_n} C_{ijmn}$$

$$\prod_{k_\mu,q_\nu} \sum_{r_{\mu,12},\bar{r}_{\mu,12}} \sum_{\rho_{\mu,12},\bar{\rho}_{\mu,12}} \{\exp[i(k_\mu r_{\mu,12} + \bar{k}_\mu \bar{r}_{\mu,12} + q_\nu \rho_{\nu,34} + \bar{q}_\nu \bar{\rho}_{\nu,34})] c^+_{r_{\mu,12}} c^+_{\bar{r}_{\mu,12}} c^+_{\rho_{\nu,34}} c^+_{\bar{\rho}_{\nu,34}} C^+_{\pm k\uparrow\downarrow} + TR_B\}$$

(4a)

$$= \sum_{i,m=0}^{N/2} (\frac{u_k}{v_k})^i (\frac{u_q}{v_q})^m \sum_{b_i,b_j,a_m,a_n} C_{ijmn} \psi_{ij,mn} \qquad (4b)$$

where $TR_B$ means the time reversal of the former term, in phase B.

$$\psi_{ij,mn} = \sum_{r_{\mu 12} \bar{r}_{\mu 12} \rho_{v34} \bar{\rho}_{v34}} \{\exp[i\sum_{\mu,v}(k_\mu r_{\mu,12} + \bar{k}_\mu \bar{r}_{\mu,12} + q_v \rho_{v,34} + \bar{q}_v \bar{\rho}_{v,34})]\prod_{\mu,v} c^+_{r_{\mu,12}} c^+_{\bar{r}_{\mu,12}} c^+_{\rho_{v,34}} c^+_{\bar{\rho}_{v,34}} C^+_{\pm k\uparrow\downarrow}\} + TR_B$$

$$= \sum_{r_{\mu 12} \bar{r}_{\mu 12} \rho_{v34} \bar{\rho}_{v34}} \{\cos[\sum_{\mu,v}(k_\mu r_{\mu,12} + \bar{k}_\mu \bar{r}_{\mu,12} + q_v \rho_{v,34} + \bar{q}_v \bar{\rho}_{v,34})]$$

$$\times[(\prod_{\mu,v} c^+_{r_{\mu,12}} c^+_{\bar{r}_{\mu,12}} c^+_{\rho_{v,34}} c^+_{\bar{\rho}_{v,34}} C^+_{\pm k\uparrow\downarrow})_A + (\prod_{\mu,v} c^+_{r_{\mu,12}} c^+_{\bar{r}_{\mu,12}} c^+_{\rho_{v,34}} c^+_{\bar{\rho}_{v,34}} C^+_{\pm k\uparrow\downarrow})_B]$$

$$+i\sin[\sum_{\mu,v}(k_\mu r_{\mu,12} + \bar{k}_\mu \bar{r}_{\mu,12} + q_v \rho_{v,34} + \bar{q}_v \bar{\rho}_{v,34})]$$

$$\times[(\prod_{\mu,v} c^+_{r_{\mu,12}} c^+_{\bar{r}_{\mu,12}} c^+_{\rho_{v,34}} c^+_{\bar{\rho}_{v,34}} C^+_{\pm k\uparrow\downarrow})_A - (\prod_{\mu,v} c^+_{r_{\mu,12}} c^+_{\bar{r}_{\mu,12}} c^+_{\rho_{v,34}} c^+_{\bar{\rho}_{v,34}} C^+_{\pm k\uparrow\downarrow})_B]\} \quad (5a)$$

$$\psi_{ij,mn} = \sum_{r_{\mu 12} \bar{r}_{\mu 12} \rho_{v34} \bar{\rho}_{v34}} \{\cos[\sum_{\mu,v}(k_\mu r_{\mu,12} + \bar{k}_\mu \bar{r}_{\mu,12} + q_v \rho_{v,34} + \bar{q}_v \bar{\rho}_{v,34})][C^A_{12,34} + C^B_{12,34}]$$

$$+i\sin[\sum_{\mu,v}(k_\mu r_{\mu,12} + \bar{k}_\mu \bar{r}_{\mu,12} + q_v \rho_{v,34} + \bar{q}_v \bar{\rho}_{v,34})][C^A_{12,34} - C^B_{12,34}]\} \quad (5b)$$

$C^+_{\pm k\uparrow\downarrow}$ is defined in [2], and the definitions of $C^A_{12,34}$ and $C^B_{12,34}$ are obvious from a comparison between Eq. (5a), and Eq. (5b). We also define $C^+_{12,34} = C^A_{12,34} + C^B_{12,34}$, and $C^-_{12,34} = C^A_{12,34} - C^B_{12,34}$.

The operators $C^\pm_{12,34}$ are defined as a sum or a difference between the operators $C^A_{12,34}$ and $C^B_{12,34}$. Each of the two later operators contains spin projections along the x-y plane. In the LSCO family the spins are aligned 45° from the Cu-O bonds [10,11], whereas in the YBCO family they are parallel to the bonds [10,12]. We assume spins along the $\pm x$ directions, without reducing the generality of the treatment. Thus, the spin function along the x direction is $\begin{bmatrix}1\\1\end{bmatrix}$, and along the $-x$ direction is $\begin{bmatrix}1\\-1\end{bmatrix}$. Recalling that each spin in phase A is the reversal of that in phase B, we get for some check points in $C^+_{12,34}$ spins in the z direction, and for their respective points in $C^-_{12,34}$ spins in the $-z$ direction. These spin directions are symmetric to rotations around an axis normal to the xy plane (except for phase changes). Thus,

$$|\psi_k \psi_q\rangle = \sum_{i,m=0}^{N/2} (\frac{u_k}{v_k})^i (\frac{u_q}{v_q})^m \sum_{b_i,b_j,a_m,a_n} C_{ijmn} \sum_{r_{\mu 12} \bar{r}_{\mu 12} \rho_{\nu 34} \bar{\rho}_{\nu 34}} \{\cos[\sum_{\mu,\nu}(k_\mu r_{\mu,12} + \bar{k}_\mu \bar{r}_{\mu,12} + q_\nu \rho_{\nu,34} + \bar{q}_\nu \bar{\rho}_{\nu,34})] C^+_{12,34}$$

$$+ i\sin[\sum_{\mu,\nu}(k_\mu r_{\mu,12} + \bar{k}_\mu \bar{r}_{\mu,12} + q_\nu \rho_{\nu,34} + \bar{q}_\nu \bar{\rho}_{\nu,34})] C^-_{12,34}\} \quad (6)$$

Now we wish to examine the symmetry of the ground state function $|\psi_0\rangle = \prod_{k,q} |\psi_k \psi_q\rangle$ to rotations by $n\pi/2$ around axes normal to the xy plane. We notice the symmetry between the transversal components of the wavevectors of rows and columns in Eq. (6). The symmetry between the longitudinal components, however, is not realized unless we examine a product of two row wavevectors with two column wavevectors, with identical longitudinal components. The ground state $|\psi_0\rangle = \prod_{k,q} |\psi_k \psi_q\rangle$ includes such products, but to avoid mathematical cumbersomeness we continue the analysis with the function of Eq. (6), without writing explicitly the longitudinal components of the wavevectors. One sees immediately that the function of Eq. (6) is symmetric to rotations by $n\pi/2$ around axes normal to the xy plane, except for the phase changes that result from the spin rotations. This is so because the summation in Eq. (6) assures that each term has also its rotation symmetric term, and also because the spins in $C^+_{12,34}$, and in $C^-_{12,34}$ are all aligned normal to the xy plane. These spins undergo only phase changes because of the rotation. One might think, that the existence of AFM modulated waves according to [1], suggests that the $C^+_{\pm k\uparrow\downarrow}$ parts in $C^+_{12,34}$ and in $C^-_{12,34}$ include spins that are aligned along the xy plane. In the next section, however, we show that this is not the case in the superconductive state, in accordance with the observation of the spin gap phenomenon. Besides, it is shown below that the final version of the superconductive state is obtained by a linear combination of the function in Eq. (6), after rotation transformations. Although (the $C^+_{\pm k\uparrow\downarrow}$ parts of) the function in Eq. (6) may include spins that are aligned along the xy plane, the described linear combination does not include such spins.

The symmetry of Eq. (6) to rotations around axes normal to the xy plane does not make it compatible with experiments. Josephson tunneling experiments suggest that pairing state function changes phase as $\Delta\phi = 2\alpha$, when $\alpha$ is the angle of rotation around an axis normal to the xy plane. This is compatible with the $d_{x^2-y^2}$ symmetry. We can make the state function compatible with experiment by making a (normalized) linear combination of the function of Eq. (6). This does not violate any quantum mechanical rule since one has the freedom to choose phases of wavefunctions. The components of the linear combination are defined as the following rotated function of Eq. (6). Let us define $R(\alpha)$ to be a rotation operator by the angle $\alpha$ around an axis normal to the xy plane. Then the following function

$$|\psi_k \psi_q> = C \sum_{i,m=0}^{N/2} \left|\frac{u_k}{v_k}\right|^i \left|\frac{u_q}{v_q}\right|^m \{ \sum_{\alpha=0,\pi} R(\alpha) \sum_{b_i,b_j,a_m,a_n} C_{ijmn}\psi_{ij,mn} - \sum_{\alpha=\pm\pi/2} R(\alpha) \sum_{b_i,b_j,a_m,a_n} C_{ijmn}\psi_{ij,mn} \}$$

, (7)

where C is a constant, has the desirable $d_{x^2-y^2}$ symmetry. This proves that our double correlated model of linear aggregates of holes is in agreement with experiment. An essential condition for this agreement is the rejection of the stripe geometry, and the assumption of the checkerboard geometry [1,2]. This is so because we cannot envision any mechanism which relates the phases of two different and perpendicular stripe domains (especially when the size of the stripe is larger than the coherence length). This inconceivability is still valid for models in which perpendicular stripes are suggested for neighboring $CuO_2$ in the planes in the z-directions [10]. Note that this argument against the stripe model adds to the one that stems from the shape of the "Fermi surface" that has been presented in [1]. Besides, within each domain the stripe model fits either the symmetry group $C_2$, or $C_{2v}$ (after ignoring small structural tilts, and incommensurability). None of these groups has a representation suitable for the symmetric transformation of $x^2 - y^2$. However, the symmetry groups $C_4$, and $C_{4v}$, that fit the checkerboard geometry, have such representations [13]. Finally we notice that the $d_{x^2-y^2}$ symmetry here has the unit cell of the incommensurate magnetic structure, namely: $a(\delta^{-1},\delta^{-1})$. Thus, the $d_{x^2-y^2}$ symmetry here should not be interpreted as the known atomic symmetry.

## 3. **THE SPIN GAP**

There is sufficient experimental evidence for superconductive energy gaps in the dispersions of the Neutron scattered peaks from cuprates, at the wavevectors: $2\pi a^{-1}(0.5\pm\delta,0.5)$, and $2\pi a^{-1}(0.5,0.5\pm\delta)$ [7-10]. In the superconductive state, these AFM modulation peaks are absent at energies below the superconductive gap, but present above it. In the normal state, the peaks were observed down to zero energy. Proving the spin gap may be a critical test to our model. In the following we show that our model is in agreement with the experimental observations of the spin gap.

The magnetic structure factor that yields the discussed AFM modulation peaks was calculated in [1] from $<\psi_0|C_j^+ C_j|\psi_0>$, where $C_j^+$ is the operator that creates a column of holes at the column index $j$, while reversing all the spins in columns $j' > j$. In the two dimensional presentation the same quantity could be calculated by means of

$$<\psi_0|\tilde{\psi}^+(x=ja)\tilde{\psi}(x=ja)|\psi_0>=$$
$$=N^{-1}<\psi_0|\sum_{|k|<2k_F} C_k^+ C_k + C_{\bar{k}}^\pm C_{\bar{k}}^\mp + [C_{\bar{k}}^\pm C_k e^{-i(\bar{k}-k)x} + C_k^+ C_{\bar{k}} e^{i(\bar{k}-k)x}]|\psi_0> \qquad (8)$$

Eq. (8), with the proper normalization constant, yields the results of [1]. When both phases A and B are taken into account their CDW interfere destructively, whereas the SDW interfere constructively. This immediately suggests a way of extension to the four dimensional presentation, which is suitable to treat superconductivity. Thus, for the energy and momentum dependence (of the magnetic waves) one should calculate $\sum_i f_i(T)<\psi_i|\prod_s \tilde{\psi}_s^+(x,t)\tilde{\psi}_s(x,t)|\psi_i>$, where $|\psi_i>$ are the excited and ground states, $f_i(T)$ are their temperature dependent weighting coefficients, and $\tilde{\psi}_s(x,t)$ is the 4-dimensional vector given by the forthcoming Eq. (17). However, in the present analysis we intend to calculate only the expectation value of $\prod_s \tilde{\psi}_s^+(x,t)\tilde{\psi}_s(x,t)$ in the ground state, and to demonstrate only the sole existence of the spin gap. Consequently, we avoid the time dependence of the field operators, and calculate only $<\psi_0|\prod_s \tilde{\psi}_s^+(x)\tilde{\psi}_s(x)|\psi_0>$. At first glance one evaluates that the results of the first and third components of $\tilde{\psi}_s(x)$ should yield the same result as the 2-dimensional one, whereas the results of the second and fourth components should have a reversed sign, and cancel out the first. However, this is only an intuitive speculation since the superconductive ground state has a different symmetry. Thus, one does not know in advance the effect of the superconductive term in $|\psi_0>$, even when one assumes that the complete columns and rows of holes are undisturbed by superconductivity. On top of that, we know that superconductivity disturbs the completeness of the columns (rows) of holes [2], and this should also be taken into account. Consequently, the calculation must be carried out in detail. In the following, first we use full column (row) operators in the basic calculation, and then show that their fluctuations should not change the results. When attempting calculations on the superconductive field

$$\tilde{\psi}_{k,s} = \begin{pmatrix} C_{k,s} \\ C^+_{-\bar{k},-s} \\ C_{\bar{k},s} \\ C^+_{-k,-s} \end{pmatrix}, \tag{9a}$$

we realize that there are also the equivalent fields $\tilde{\psi}_{k,-s}$, $\tilde{\psi}_{-k,s}$, and $\tilde{\psi}_{-k,-s}$. The fields $\tilde{\psi}_k$ and $\tilde{\psi}(x)$ are defined on a full column of holes and includes both spins,

$$\tilde{\psi}_k(x) = \tilde{\psi}_{k,s}\tilde{\psi}_{k,-s} = \begin{pmatrix} C_k e^{ikx} \\ C^+_{-\bar{k}} e^{i\bar{k}x} \\ C_{\bar{k}} e^{i\bar{k}x} \\ C^+_{-k} e^{ikx} \end{pmatrix} \tag{9b}$$

$$\tilde{\psi}(x) = \frac{1}{2}\sum_k \begin{pmatrix} C_k e^{ikx} \\ C^+_{-\bar{k}} e^{i\bar{k}x} \\ C_{\bar{k}} e^{i\bar{k}x} \\ C^+_{-k} e^{ikx} \end{pmatrix} \qquad (9c)$$

In Eqs. (9b) and (9c) the $C_k$ operators are full column (row) hole operators (including two spin states). The x dependence in Eq. (9b) is based on the fact that the $C_{k,s}$ and $Ck,-s$ operators have the same x parameter. Now we get

$$\tilde{\psi}^+(x)\tilde{\psi}(x) =$$

$$= \sum_{|k|<k_F} [C_k^+ C_k + C_{\bar{k}}^+ C_{\bar{k}} + C_{\bar{k}}^+ C_k(x) + C_k^+ C_{\bar{k}}(x) + 2 - C_{-k}^+ C_{-k} - C_{-\bar{k}}^+ C_{-\bar{k}} - C_{-\bar{k}}^+ C_{-k}(x) - C_{-k}^+ C_{-\bar{k}}(x)]$$

$$(10)$$

The ground state is

$$|\psi_0> = \prod_{|q|<k_F} [\tilde{v}_q C_q^+ C_{-q}^+ + \tilde{w}_q (C_q^+ C_{-\bar{q}}^+ + C_{\bar{q}}^+ C_{-q}^+) + \tilde{\theta}_q C_{\bar{q}}^+ C_{-\bar{q}}^+ + \tilde{u}_q (1 + C_q^+ C_{-q}^+ C_{\bar{q}}^+ C_{-\bar{q}}^+)]|0>$$

$$(11)$$

Here full Column operators are used, in contrast with Eq. (15) in [2], where the superconductive part is a sum of weighted combinations with different number of holes. This simplification, though, does not change the consequences, since the application of $\tilde{\psi}^+(x)\tilde{\psi}(x)$ on each superconductive (combinations) term vanishes separately. A straightforward calculation of $\tilde{\psi}^+(x)\tilde{\psi}(x)|\psi_0>$ shows that the last four terms out cancel the first four terms, so that $<\psi_0|\tilde{\psi}^+(x)\tilde{\psi}(x)|\psi_0> = const$, independent of x. This, in contrast to Eq. (8), proves the phenomenon of the spin gap at zero temperature. At $T > T_c$, the superconductive symmetry break down, the two dimensional treatment of Eq. (8) is valid, and the magnetic peaks are observable.

It should be stressed out that the cancellation that is caused by the two parts of Eq. (10) should not be considered similar to the cancellation of the CDW of the two phases A and B in Eq. (27) of [1]. The elimination of the CDW there implies a constructive interference of the SDW's of the two different phases. Here the two parts in Eq. (10) that have different signs belong to the same magnetic phase, as Eqs. (8-10) were defined on one magnetic phase. This is so despite that the components of $\tilde{\psi}_{k,s}$ are seemingly the time reversals of each other (the fourth of the first, and the second of the third). In fact, the various components of $\tilde{\psi}_{k,s}$ are not time reversal of each other because of the background spins. The field $\tilde{\psi}_{k,s}$ is defined separately on each different magnetic phase, A and B, and the same calculations could be defined on the other phase, which does not change the consequences about the spin-gap.

We also notice that there is a way to conclude the existence of the spin gap from symmetry of the superconductive function of Eq. (7), which is composed of transformed parts of the function of Eq. (6), so that the spin parts that were included in $C^+_{\pm k \uparrow \downarrow}$ - are neutralized.

## 4. **THE UNPERTURBED HAMILTONIAN**

In this section I wish to re-derive the Hamiltonian- $H_0$ and the propagator $G_0$, but before doing so let us reaffirm a basic characteristic of the column (row) annihilation and creation operators- $C_j$, and $C_j^+$. In Ref. [1], these operators were considered to have Fermionic anti-commutation relations, based on the arbitrary assumption that they are made of an odd number of holes. Some of the readers may feel an objection to this assumption because of the following reasons. 1) The symmetry between the two spin states suggests an even number of holes in a column (row). 2) A column (row) of holes "consumes" two columns (rows) of states in the antinodal sections of the "Fermi area" in the reciprocal space, which might suggest an even number of holes. Here however, we shall reaffirm the Fermionic characteristics of the columns (rows) based on a different argument. Let us for the sake of simplicity concentrate only on columns. We define $|0>$ to be the vacuum of columns, so that $C_j|0>=0$, and $C_j C_j^+|0>=1|0>$, and $C_j^+ C_j^+|0>=0$, for any $j$. These relations are obvious because of the basic physics of the cuprates, where one deals only with planar $CuO_2$, where the $d_{x^2-y^2}$ states of the Cu are $\sigma$ anti-bonded to the p states of Oxygen. The anti-ferromagnetic state $|0>$ is obtained when the number of spins is equal to the number of sites, and $C_j^+|0>$ is obtained when N holes are aggregated to form a column. The creation of another string of holes on the same column index is impossible, as the elimination of a string of holes where it does not exist. All these suggest the Fermionic relations $\{C_i, C_j^+\} = \delta_{i,j}$, and $\{C_i, C_j\} = 0$, for every $i,j$. The suitable relations in the wavevector space follow immediately.

The unperturbed Hamiltonian $H_0$ was derived in [1,2]. Unfortunately, the 4x4 matrix Hamiltonian in [2] does not match the 2x2 Hamiltonian in [1], when the superconductive parameter $u_k$ is set equal to zero. The disagreement has to do with the spatial dependent part in the 4-dimensional case. In Eq. (30b) of [2] it shows in the diagonal part as: $-\tau_3 F_k(x)$. Examination of the derivation procedure reveals the reason. Each of the elements of the vector field $O_k$ has two different wavevectors associated to its components, and therefore, it is incorrect to Fourier transform $O_k$ by the single factor: $\exp(ikx)$. The correct procedure is the one exercised in Eqs. (11) and (12a,b) in [2], where $c_k$ and $c_{\bar{k}}$ were Fourier associated to their wavevectors. Consequently, we redo the derivations of $H_0$ in 2 and 4 dimensions by this procedure. In 2-dimension the field in phase A is

$$\tilde{\psi}_{kA}(t) = \begin{pmatrix} -w_k \\ v_k \end{pmatrix} \gamma_{kA} e^{-iE_k t} + \begin{pmatrix} v_k \\ w_k \end{pmatrix} \eta_{kA}^+ e^{iE_k t} = (M_k e^{-iE_k t} + N_k e^{iE_k t}) \begin{pmatrix} C_k \\ C_{\bar{k}} \end{pmatrix}_A \quad (12)$$

Here the operators are column operators, and

$$M_k = \begin{pmatrix} w_k^2 & -w_k v_k \\ -w_k v_k & v_k^2 \end{pmatrix} \quad (13a)$$

$$N_k = \begin{pmatrix} v_k^2 & w_k v_k \\ w_k v_k & w_k^2 \end{pmatrix} \quad (13b)$$

Consequently

$$H_{0,A} = \frac{1}{2}\sum_k E_k \begin{pmatrix} C_k^+ e^{-ikx} & C_{\bar{k}}^+ e^{-i\bar{k}x} \end{pmatrix}_A (M_k - N_k) \begin{pmatrix} C_k e^{ikx} \\ C_{\bar{k}} e^{i\bar{k}x} \end{pmatrix}_A$$

$$+ \frac{1}{2}\sum_{k,k'} E_k \delta_{k'\bar{k}} \begin{pmatrix} C_{k'}^+ e^{-ik'x} & C_{\bar{k}'}^+ e^{-i\bar{k}'x} \end{pmatrix}_A (M_{k'} M_k - N_{k'} N_k) \begin{pmatrix} C_k e^{ikx} \\ C_{\bar{k}} e^{i\bar{k}x} \end{pmatrix}_A . \quad (14)$$

The sums are over wavevectors range that doubles the Fermi range. The reason for the pre-factor 1/2 is that, in the said wavevector range, each operator appears twice. This pre-factor should become unity if the wavenumber sum is only for $|k|<k_F$. Since: $M_{\bar{k}} M_k - N_{\bar{k}} N_k = -2 w_k v_k \tau_1$, we get

$$H_{0,A} = \sum_{|k|<k_F} \{\tilde{\psi}_{k,A}^+ (\varepsilon_k \tau_3 + \Lambda_{k,A} I) \tilde{\psi}_{k,A} + \tilde{\psi}_{k,A}^+(x)(\Lambda_{k,A}\tau_1)\tilde{\psi}_{k,A}(x)\} \quad (15a)$$

where $\varepsilon_k = E_k(w^2 - v^2)$, and $\Lambda_k = -2w_k v_k E_k$. The Hamiltonian of Eq. (15a) applies only for the functions of the magnetic phase A. For phase B the parameter $w_k$ changes sign in all the relevant equations -( Eqs. (12)-(15)). Diagonalization of $H_{0,A}$ yields

$$H_{0,A} = \sum_{|k|<k_F} \begin{pmatrix} \gamma_{kA}^+ & \eta_{kA} \end{pmatrix} \begin{pmatrix} E_k + \Lambda_{k,A} & 0 \\ 0 & -E_k + \Lambda_{k,A} \end{pmatrix} \begin{pmatrix} \gamma_{kA} \\ \eta_{kA}^+ \end{pmatrix} \quad (15b)$$

We see that the obtained eigenvalues are inconsistent with the initially assumed excitation energies- $E_k$, in Eq. (8). The difference is the additional term $I\Lambda_{k,A}$, in Eq. (15b). This inconsistency is removed when one recalculates the Hamiltonian with the assumption of the energy eigenvalues of Eq. (15b), but with wavevector summation only for $|k|<k_F$. Such a procedure removes the inconsistency, and yield back Eqs. (15a,b).

The 4-dimensional field is given by Eqs. (9). Two such fields are defined, one for the phase A, and one for B. The space and time dependent field is defined in accordance with the basic excitations of $O_k(t)$ in [2]

$$\tilde{\psi}_s(x,t) = \sum_k (M_k e^{-iE_k t} + N_k e^{iE_k t}) \begin{pmatrix} C_{k,s} e^{ikx} \\ C^+_{-\bar{k},-s} e^{i\bar{k}x} \\ C_{\bar{k},s} e^{i\bar{k}x} \\ C^+_{-k,-s} e^{ikx} \end{pmatrix}. \tag{16}$$

In Eq. (16), the subscript s denotes the spin state, which has two values for each wavevector. The field operators $\tilde{\psi}_{k,s}$ are also defined for both spin states for each wavevector (despite of the appearance of $C^+_{-k,-s}$ in the definition of $\tilde{\psi}_{k,s}$). Recalling that $C_k$ is a column operator that includes both spins, $C_{k,s}$ is the part in it that relates only to the spin projection s, as defined in Eqs. (8) in [2]. We also have,

$$M_k = -u_k \alpha_1 - w_k \alpha_3 + \frac{1}{2} I + \frac{1}{2}(\theta_k - v_k)\beta. \tag{17a}$$

$$N_k = u_k \alpha_1 + w_k \alpha_3 + \frac{1}{2} I - \frac{1}{2}(\theta_k - v_k)\beta. \tag{17b}$$

As for the 2-dimensions, we have $M_k^2 = M_k$, and $N_k^2 = N_k$. The unperturbed Hamiltonian then should be

$$H_0 = \frac{1}{2} \sum_k E_k \begin{pmatrix} C^+_{k,s} e^{-ikx} & C_{-\bar{k},-s} e^{-i\bar{k}x} & C^+_{\bar{k},s} e^{-i\bar{k}x} & C_{-k,-s} e^{-ikx} \end{pmatrix} (M_k - N_k) \begin{pmatrix} C_{k,s} e^{ikx} \\ C^+_{-\bar{k},-s} e^{i\bar{k}x} \\ C_{\bar{k},s} e^{i\bar{k}x} \\ C^+_{-k,-s} e^{ikx} \end{pmatrix}$$

$$+\frac{1}{2}\sum_k E_k \begin{pmatrix} C^+_{\bar{k},s}e^{-i\bar{k}x} & C_{-k,-s}e^{-ikx} & C^+_{k,s}e^{-ikx} & C_{-\bar{k},-s}e^{-i\bar{k}x} \end{pmatrix}(M_{\bar{k}}M_k - N_{\bar{k}}N_k)\begin{pmatrix} C_{k,s}e^{ikx} \\ C^+_{-\bar{k},-s}e^{i\bar{k}x} \\ C_{\bar{k},s}e^{i\bar{k}x} \\ C^+_{-k,-s}e^{ikx} \end{pmatrix} \quad (18)$$

One immediately finds that $M_{\bar{k}}M_k = -(2w_k\alpha_3 + 2u_k\alpha_1)M_k$, $N_{\bar{k}}N_k = (2w_k\alpha_3 + 2u_k\alpha_1)N_k$, so that $M_{\bar{k}}M_k - N_{\bar{k}}N_k = -2w_k\alpha_3 - 2u_k\alpha_1$. Thus, we have

$$H_0 = \sum_{|k|<k_F} \tilde{\psi}^+_{k,s}(x)\left(\varepsilon_k\beta + \Lambda_k\tau_3 + \Lambda_k\alpha_3 + \Delta_k\alpha_1 + \Delta_k\tau_1\right)\tilde{\psi}_{k,s}(x) \quad (19)$$

In Eq. (19), $\varepsilon_k = E_k(\theta_k - v_k)$, $\Lambda_{k,A} = -2w_k E_k$, $\Lambda_{k,B} = 2w_k E_k$, $\Delta_k = -2u_k E_k$, and $\tilde{\psi}_k(x)$ means that each component of $\tilde{\psi}_k$ is multiplied by its Fourier exponent. This makes only the $\alpha_3$, and the $\tau_1$ components of the Hamiltonian to be x-dependent. The Hamiltonian of Eq. (20a) is in agreement with the two dimensional Hamiltonian of Eq. (15a), and is in **disagreement** with Eq. (30b) of [2]. The agreement with Eq. (15a) is easily realized when the second and third rows and columns of the Hamiltonian are exchanged

$$H_0 = \sum_{|k|<k_F} \begin{pmatrix} C^+_{k,s} & C^+_{\bar{k},s} & C_{-\bar{k},-s} & C_{-k,-s} \end{pmatrix} \begin{pmatrix} \varepsilon_k + \Lambda_k & \Lambda_k & \Delta_k & \Delta_k \\ \Lambda_k & -\varepsilon_k + \Lambda_k & \Delta_k & \Delta_k \\ \Delta_k & \Delta_k & \varepsilon_k - \Lambda_k & -\Lambda_k \\ \Delta_k & \Delta_k & -\Lambda_k & -\varepsilon_k - \Lambda_k \end{pmatrix} \begin{pmatrix} C_{k,s} \\ C_{\bar{k},s} \\ C^+_{-\bar{k},-s} \\ C^+_{-k,-s} \end{pmatrix} \quad (20)$$

In Eq. (20), the x-dependence of the field operators is not shown explicitly. By solving the characteristic equation, the eigenvalues are found to be

$$\lambda^2_{k\pm} = [\sqrt{\varepsilon_k^2 + (\Lambda_k^2 + \Delta_k^2)} \pm \sqrt{\Lambda_k^2 + \Delta_k^2}]^2 = [E_k \pm F_k]^2. \quad (21)$$

Note that here $F_k$, and $E_k$, are independent on x. The diagonal Hamiltonian is

$$H_{0A} = \sum_{|k|<k_F} \begin{pmatrix} \eta_k^+ & \gamma_k^+ & \rho_k & \sigma_k \end{pmatrix} (\beta E_k - \tau_3 F_k) \begin{pmatrix} \eta_k \\ \gamma_k \\ \rho_k^+ \\ \sigma_k^+ \end{pmatrix} \qquad (22)$$

For $H_{0B}$, the sign of $\tau_3 F_k$ is reversed.

We notice from Eq. (15b) that the energy scales of the magnetic phases A and B are shifted by $\pm|\Lambda_k|$, where the minus (plus) sign applies to the phase A (B), respectively. This suggests that pair excitations, such as $\eta_{k_F-\delta,B}^+ \gamma_{k_F+\delta,A}^+$, require only infinitesimally small energy, when $\delta$ is infinitesimally small. This inspired the proposed zero gap field in [1]

$$\Psi_{k-}(t) = \begin{pmatrix} -w_k \\ v_k \end{pmatrix} \gamma_{kA} \exp(-iE_{k-}t) + \begin{pmatrix} v_k \\ -w_k \end{pmatrix} \eta_{kB}^+ \exp(iE_{k-}t), \qquad (23)$$

where $E_{k-} = \sqrt{\varepsilon_k^2 + \Lambda_k^2} - |\Lambda_k|$. The propagator appropriate to this field is [1]

$$G_0(k,\omega) = \frac{1}{2}[\frac{I+(w_k^2-v_k^2)\tau_3 - 2w_k v_k \tau_1}{\omega - E_{k-} + i\delta\omega} + \frac{I-(w_k^2-v_k^2)\tau_3 - 2w_k v_k \tau_1}{\omega + E_{k-} - i\delta\omega}] \ . \qquad (24)$$

In Eq. (24), the first term applies to $\gamma_{kA}^+$ excitations, whereas the second term applies to $\eta_{kB}^+$ excitations.

## 5. **CONCLUSIONS**

Two important features of the superconductive state of the cuprates have been demonstrated in the present paper, the pairing symmetry, and the "spin gap". The treatment demonstrates that our model is not only intuitively compatible with the experimental reality, but may also be rigorously proven to be in agreement with it. This is also the situation with the former comparison with the "Fermi surfaces" obtained by ARPES experiments, and with the Neutron scattering, and STM results. All these comparisons with experiments are based upon semi-one-dimensional symmetry breaking that is originated from linear aggregations of holes in checkerboard geometry. The aggregated holes undergo correlations of two types, one that is responsible for the pseudogap, and the other that is responsible for the superconductivity. In the present paper the demonstration of the spin gap is has resulted from the definition of the field in the superconductive state, as evident from Eq. (10). It is also related to the pair symmetry, as is evident from the discussion following Eq. (7).

We believe that the model has not yet of full maturity. There are still developments, modifications, and adjustments to be done. We also believe, however, that its basic principles and structure are correct.